\documentclass[aps, prd, showpacs, showkeys, preprint, eqsecnum, floatfix, superscriptaddress, nofootinbib, longbibliography]{revtex4-2}

\usepackage{hyperref}
\usepackage{amssymb}
\usepackage{amsmath}
\usepackage{amsthm}
\usepackage{bm}
\usepackage{calrsfs}
\usepackage{color}
\usepackage{graphicx}
\usepackage[normalem]{ulem} 
\usepackage{relsize}
\usepackage{multirow}

\newcommand{\bq}{\begin{eqnarray}}
\newcommand{\eq}{\end{eqnarray}}
\newcommand{\bqn}{\begin{eqnarray*}}
\newcommand{\eqn}{\end{eqnarray*}}

\newcommand{\ZZ}{{\rm Z}\hskip-.3em{\rm Z}}

\newcommand{\calo}{{\cal O}}

\def\bk{\mathbf k}
\def\bl{\mathbf l}

\begin{document}
%%%%%%%%%%%%%%%%%%%%%%%%%%%%%%%%%%%%%%%%%%%%%%%%%%%%%%%%%%%%%%%%%%%%%%%%%%%%%%
%%%%%%%%%%%%%%%%%%%%%%%%%%%%%%%%%%%%%%%%%%%%%%%%%%%%%%%%%%%%%%%%%%%%%%%%%%%%%%
%%%%%%%%%%%%%%%%%%%%%%%%%%%%%%%%%%%%%%%%%%%%%%%%%%%%%%%%%%%%%%%%%%%%%%%%%%%%%%
\title{Monte Carlo evaluation of the continuum limit of the two-point function of the  
Euclidean free real scalar field subject to affine quantization} 

\author{Riccardo Fantoni}
\email{riccardo.fantoni@posta.istruzione.it}
\affiliation{Universit\`a di Trieste, Dipartimento di Fisica, strada
  Costiera 11, 34151 Grignano (Trieste), Italy}

%\author{Heinrich Leutwyler}
%\email{Leutwyler@itp.unibe.ch}
%\affiliation{}

\author{John R. Klauder}
\email{klauder@phys.ufl.edu}
\affiliation{Department of Physics and Department of Mathematics \\
University of Florida,   %P.O. Box 118440\\
Gainesville, FL 32611-8440}

%\author{Laure Gouba}
%\email{lgouba@ictp.it}
%\affiliation{Abdus Salam International Centre for Theoretical Physics, ICTP
%Strada Costiera, 11, I-34151 Trieste Italy}
\date{\today}

\begin{abstract}
We study canonical and affine versions of the quantized covariant Euclidean free real 
scalar field-theory on four dimensional lattices through the Monte Carlo method. We 
calculate the two-point function near the continuum limit at finite volume. Our 
investigation shows that affine quantization is able to give meaningful results for the 
two-point function for which is not available an exact analytic result and therefore 
numerical methods are necessary. 
\end{abstract}



\maketitle
%%%%%%%%%%%%%%%%%%%%%%%%%%%%%%%%%%%%%%%%%%%%%%%%%%%%%%%%%%%%%%%%%%%%%%%%%%%%%%
\section{Introduction}
%%%%%%%%%%%%%%%%%%%%%%%%%%%%%%%%%%%%%%%%%%%%%%%%%%%%%%%%%%%%%%%%%%%%%%%%%%%%%%
\label{sec:introduction}

The aim of this work is to find out what affine quantization does to a classical 
field-theory. The simplest such theory is a free real scalar field of mass $m$. In that 
case, the spectrum of physical states obtained with canonical quantization is known: 
states containing many indistinguishable particles with momenta $\vec{p}_1,\vec{p}_2,
\ldots$ and energies $\sqrt{|\vec{p}_i|^2+m^2}$ (here $c=1$) obeying Bose statistics. 
The simplest question to ask now is: what becomes of this if the free real scalar field 
is subject to affine quantization \cite{Klauder2020c,Klauder2000} rather than 
canonical quantization \citep{Dirac}? Does the system describe particles in this case 
as well? If so, do they interact with one another? Working out the two-point function 
of the free field in that framework should be of use to answer these questions. 

The free real scalar field is well understood by canonical quantization.
The standard set of problems that can be resolved by canonical quantization
is distinct from the standard set of problems that can be resolved by affine
quantization, and one can therefore expect that an affine quantization of
the classical free real scalar differs from that of canonical quantization. The
purpose of this paper is to try to understand in what ways an affine quantization
is similar as well as dissimilar from a canonical quantization. We add
that some non-free real scalar fields have already been observed and that canonical
quantization fails for several non-renormalizable fields, such as $(\phi^{12})_3$ 
\cite{Fantoni2020} and $(\phi^4)_4$ \cite{Fantoni2020a}. The key to that result
is the introduction of a highly unusual, additional, non-quadratic, term
that is dictated by affine quantization. While affine quantization employs
an additional term, that particular term formally disappears when $\hbar\to 0$,
which makes it a plausible modification of the quadratic terms of traditional
free real scalar fields in order to extend acceptable quantization of traditional
non–renormalizable models.

The Euclidean action in canonical quantization \citep{Dirac}, in units where $\hbar=1$, 
is
\bq \label{eq:c-action}
S^{(c)}[\phi]=\mathop{\mathlarger{\mathlarger{\int}}}\left\{\frac{1}{2}
\sum_{\mu=0}^s\left[\frac{\partial\phi(x)}{\partial x_\mu}\right]^2 
+V(\phi(x))\right\}\,d^nx,
\eq
with $x=(x_0,x_1,\ldots,x_s)=(x_0,\vec{x})$ for $s$ spatial dimensions and $n=s+1$ for 
the number of space-time dimensions with $x_0=ct$. We will work at $s=3$. And $V$ is 
the self-interaction potential density for which we will choose 
$V(\phi)=(1/2)m^2\phi^2$ corresponding to a free-theory with a bare mass $m$.

The Eudlidean action in affine quantization \cite{Klauder2020c,Klauder2000} is
\bq \label{eq:a-action}
S^{(a)}[\phi]=\mathop{\mathlarger{\mathlarger{\int}}}\left\{\frac{1}{2}
\sum_{\mu=0}^s\left[\frac{\partial\phi(x)}{\partial x_\mu}\right]^2 
+\frac{3}{8}\frac{\delta^{2s}(0)}{\phi^2(x)+
\epsilon}+V(\phi(x))\right\}\,d^nx,
\eq
where $\epsilon>0$ is a parameter used to regularize the ``$3/8$'' extra term (see 
Appendix A in \cite{Fantoni2020}) and $\delta$ is a Dirac delta function. In this case 
the Hamiltonian density contains a divergent term, in the total 
potential density 
${\cal V}(\phi)=\frac{1}{2}m^2\phi^2+\frac{3}{8}\delta^s(0)/(\phi^2+\epsilon)$, in the 
continuum, but the field theory can be treated on a lattice, and the {\sl approach} 
toward the continuum will be taken under exam in this work. In fact, the path integral 
needs this feature since we have examples such as  
$\int\phi^2(x) e^{- S^{(a)}[\phi]}\,{\cal D}\phi/\int e^{-S^{(a)}[\phi]}{\cal D}\phi$
which is a creation of $\langle \psi| \hat{\phi \mkern 0mu}^2(x) |\psi\rangle$, namely 
it creates a quantum version of the classical $\phi^2(x)$. The quantum operator
$\hat{\phi \mkern 0mu}^2(x) \sim \delta^s(0)$ and must be passed through the functional 
integral which deals with terms within $S^{(a)}[\phi]$ leading to the fact that the 
term $\phi^2(x)$ needs to be $\sim \delta^s(0)$ (at the minima of ${\cal V}$) to handle 
the integration and that factor being ``passed'' to the quantum operator term 
$\hat{\phi \mkern 0mu}^2(x)$. In the $V\to 0$ limit, this model remains different from 
a massless free-theory due to exactly the new 
$(3/8)\delta^{2s}(0)/[\phi^2(x)+\epsilon]$ 
interaction term (we have a ``pseudofree'' situation).

In our previous works we studied the non-renormalizable canonical cases with
$V(\phi)=(1/2)m^2\phi^2+g\phi^4$ \cite{Fantoni2020a} in $s=3$ and $(1/2)m^2\phi^2+g
\phi^{12}$ in $s=2$ \cite{Fantoni2020}, where $g$ is the bare coupling constant. And 
we showed that the corresponding affine cases are indeed renormalizable.

Monte Carlo (MC) \cite{Kalos-Whitlock,Metropolis} is the numerical method of choice to 
treat multidimensional integrals of high $D$ dimensions (it supercedes the traditional 
integration methods, like the trapezoidal rule, the Simpson rule,$\ldots$, based on 
the knowledge of the $\alpha^{\rm th}$ derivative of the integrating function already 
for $D>2\alpha$) therefore is especially useful to compute path integrals. We will use 
it to study the two-point function of the Euclidean action of a real scalar field in 
affine quantization. Our estimate of the path integrals will be generally subject to 
three sources of numerical uncertainties: The one due to the statistical errors, the 
one due to the space-time discretization, and the one due to the finite-size effects. 
Of these the statistical errors scale like $M^{-1/2}$ where $M$ is the computer time, 
the discretization of space-time is responsible for the distance from the continuum 
limit (which corresponds to a lattice spacing $a\to 0$), and the finite-size effects 
stems from the necessity to approximate the infinite space-time system with one in a 
periodic box of volume $L^n$ with $L=Na$ being the box side, subject to $N$ 
discretization points.

The work is organized as follows: In section \ref{sec:model} we derive the lattice 
formulation of the field theory needed in the treatment on the computer, in section 
\ref{sec:observables} we describe our computer experiment and introduce the observables 
that will be measured during our simulations, in section \ref{sec:results} we present 
our results, and section \ref{sec:conclusions} is for final remarks.

%%%%%%%%%%%%%%%%%%%%%%%%%%%%%%%%%%%%%%%%%%%%%%%%%%%%%%%%%%%%%%%%%%%%%%%%%%%%%%
\section{The lattice formulation of the field-theory model}
%%%%%%%%%%%%%%%%%%%%%%%%%%%%%%%%%%%%%%%%%%%%%%%%%%%%%%%%%%%%%%%%%%%%%%%%%%%%%%
\label{sec:model}

We used a lattice formulation of the field theory. The theory considers a real 
scalar field $\phi$ taking the value $\phi(x)$ on each site of a periodic, 
hypercubic, $n$-dimensional lattice of lattice spacing $a$ and periodicity $L=Na$. 
The canonical action for the field, Eq. (\ref{eq:c-action}), is then approximated by
\bq \label{eq:pa}
S^{(c)}[\phi]\approx\left\{\frac{1}{2}\sum_{x,\mu}a^{-2}
\left[\phi(x)-\phi(x+e_\mu)\right]^2+\sum_xV(\phi(x))\right\}a^n,
\eq
where $e_\mu$ is a vector of length $a$ in the $+\mu$ direction and we are at a 
temperature $T=1/Na$, in units where Boltzmann constant $k_B=1$. An analogous 
expression holds for the affine action of Eq. (\ref{eq:a-action}) where the Dirac 
delta function is replaced by $\delta^{2s}(0)\to a^{-2s}$.

We will use this ``primitive approximation'' for the action even if it can be improved 
in several ways \citep{Ceperley1995} in order to reduce the error due to the 
space-time discretization. In reaching to the expression (\ref{eq:pa}) we neglected the 
term $\propto a^{2n}$ due to the commutator of the kinetic and potential parts of the 
Hamiltonian, in the Baker–Campbell–Hausdorff formula. In reaching to the path integral 
expression this is justified by the Trotter formula.

The affine regularization of the previous paragraphs, leading to $\vec{x}\to \bk a$, 
where $a>0$ is the tiny lattice spacing, is helpful in our analysis but needs not be 
the final regularization. In particular, the new term $\phi(x_0,\vec{x})^{-2}\to \phi_
\bk^{-2}$ leads to a divergence when, at a fixed value of $\bk$, the integral over the 
region $|\phi_\bk|<1$, of $\int (\phi_\bk)^{-2}\;d\phi_\bk =\infty$. This behavior can 
be overcome in an additional form of regularization.\footnote{The additional 
regularization is essentially taken from Eq.~(14) in \cite{Klauder2020d}.} Instead of 
just  $\phi_\bk$ we choose $2s$ additional terms that are nearest neighbors to $\bk$. 
These additional terms enter in the form  
$\phi_\bk^{-2}\to [\;\sum_\bl \,J_{\bk,\bl}\, \phi_{\bl}^2\,]^{-1}$, where 
$J_{\bk,\bl}=(2s+1)^{-1}$ for $\bl=\bk$ plus $\bl$ is each of the $2s$ nearest 
neighbors of $\bk$. This averaging of $\phi_\bk$ also leads to a finite integration 
where, with all $|\phi_\bl|<1$, we have
\bq
\int\cdots\int \;\left[\,\sum_{\bl}\,J_{\bk,\bl}\, \phi_{\bl}^2\,\right]^{-1}\;\prod_{\bl}\,d\phi_\bl<\infty,
\eq
which is finite as determined by choosing $\phi_\bl=r \,u_\bl$ such that 
$\sum_l u_\bl^2 <\infty$ leading to the integral $U\int \,r^{-2} r^{2s} dr<\infty$, for 
all $s>0$, where  $U<\infty$ accounts for the remaining finite integrations.
   
Clearly, this procedure of averaging the expression $\phi_\bk^{-2}$ offers a smoother 
regulation, and we shall also adopt that procedure for our MC studies. We will refer to 
this affine regularization as term B and the one discussed earlier, obtained by
 choosing $J_{\bk,\bl}=\delta_{\bk,\bl}$, as term A.

The vacuum expectation of a functional observable $\calo[\phi]$ is
\bq \label{eq:expectation}
\langle\calo\rangle\approx\frac{\int\calo[\phi]\exp(-S[\phi])\,\prod_{x}d\phi(x)}
{\int\exp(-S[\phi])\,\prod_{x}d\phi(x)}, 
\eq
for a given action $S$.

We will approach the continuum limit by choosing a fixed $L$ and increasing the 
number of discretizations $N$ of each component of the space-time. So that the 
lattice spacing $a=L/N\to 0$. To make contact with the continuum limit, two conditions 
must be met $a \ll 1/m \ll L$ where $1/m$ is the Compton wavelength.

%%%%%%%%%%%%%%%%%%%%%%%%%%%%%%%%%%%%%%%%%%%%%%%%%%%%%%%%%%%%%%%%%%%%%%%%%%%%%%
\section{Simulation details and Relevant observables} 
%%%%%%%%%%%%%%%%%%%%%%%%%%%%%%%%%%%%%%%%%%%%%%%%%%%%%%%%%%%%%%%%%%%%%%%%%%%%%%
\label{sec:observables}

We want to determine the two-point function
\bq \label{eq:Fxy}
K(x,y)=\langle[\phi(x)-\langle\phi(x)\rangle][\phi(y)-\langle\phi(y)\rangle]\rangle=\langle\phi(x)\phi(y)\rangle-\langle\phi(x)\rangle^2,
\eq
replacing $x$ by $x+k$ with $k=a w_n$ with $w_n=(n_0,n_1,\ldots,n_s)$ and 
$n_\mu\in\ZZ$ amounts to a mere relabeling of the lattice points. Hence, due to 
translational invariance, $K(x,y)$ can only depend on the difference between the 
coordinates of the two points and we can define,
\bq \label{eq:tp}
D(z)=\frac{1}{L^n}\sum_x K(x,x+z)a^n,
\eq

For the massless free-theory with $V\to 0$ in canonical quantization, we find that in 
non periodic space-time (at zero temperature)
\bq
D'(z)=\int \frac{e^{-ip\cdot z}}{p^2}\,\frac{d^np}{(2\pi)^n}=
\left\{\begin{array}{ll}
-|z|/2            & n=1 \\
-(\ln |z|/l)/2\pi & n=2 \\
1/|z|4\pi         & n=3 \\
1/|z|^24\pi^2     & n=4 
\end{array}\right.,
\eq
where $|z|=\sqrt{z_0^2+z_1^2+\ldots+z_s^2}$ and $l$ is a length. This shows how the 
massless field generates long range interactions. 

For a massive free-theory with $V(\phi(x))=\frac{1}{2}m^2\phi^2(x)$ in canonical 
quantization,  we find that in non periodic space-time (at zero temperature) with n=4
\bq
D'(z)=\int \frac{e^{-ip\cdot z}}{p^2+m^2}\,\frac{d^np}{(2\pi)^n}=
m{\rm K}_1(m|z|)/|z|4\pi^2,
\eq
where $m$ is the mass and ${\rm K}_1$ is a modified Bessel function.

In periodic space-time (at a temperature $T=1/Na$)
\bq \label{eq:tpt}
D(z)=\sum_{w_n} D'(z+Lw_n),
\eq
where the sum can be restricted by an infrared cutoff $irc$ such that 
$-irc\le n_\mu\le irc$ (without any physical significance) in order to reach a given 
numerical accuracy. If we remove the cutoff the function diverges for the massless 
case.  

Our MC simulations use the Metropolis algorithm \citep{Kalos-Whitlock,Metropolis} 
to calculate the ensemble average of Eq. (\ref{eq:expectation}) which is a $N^n$ 
multidimensional integral. The simulation is started from the initial condition 
$\phi=0$. One MC step consisted in a random displacement of each one of the $N^n$ 
$\phi(x)$ as follows
\bq \label{eq:move}
\phi\rightarrow\phi+(2\eta-1)\delta,
\eq
where $\eta$ is a uniform pseudo random number in $[0,1]$ and $\delta$ is the 
amplitude of the displacement. Each one of these $N^n$ moves is accepted if 
$\exp(-\Delta S)>\eta$ where $\Delta S$ is the change in the action due to the move  
(it can be efficiently calculated considering how the kinetic part and the 
potential part change by the displacement of a single $\phi(x)$)
and rejected otherwise. The amplitude $\delta$ is chosen in such a way to have 
acceptance ratios as close as possible to $1/2$ and is kept constant during the 
evolution of the simulation. One simulation consisted of 
$M$ MC steps. The statistical error on the average $\langle\calo\rangle$ will then 
depend on the correlation time necessary to decorrelate the property $\calo$, $\tau_
\calo$, and will be determined as $\sqrt{\tau_\calo\sigma_\calo^2/(MN^n)}$, where $
\sigma_\calo^2$ is the intrinsic variance for $\calo$.  

%%%%%%%%%%%%%%%%%%%%%%%%%%%%%%%%%%%%%%%%%%%%%%%%%%%%%%%%%%%%%%%%%%%%%%%%%%%%%%
\section{Simulation results} 
%%%%%%%%%%%%%%%%%%%%%%%%%%%%%%%%%%%%%%%%%%%%%%%%%%%%%%%%%%%%%%%%%%%%%%%%%%%%%%
\label{sec:results}

We worked in units where $c=\hbar=k_B=1$. We chose the regularization parameter of 
the affine quantization A term to be $\epsilon=10^{-10}$. 
\footnote{Note that we could as well choose a regularization putting hard walls at 
$\phi=\pm\varepsilon$ therefore rejecting MC moves whenever 
$\phi\in[-\varepsilon,\varepsilon]$}

\begin{figure}[htbp]
\begin{center}
\includegraphics[width=10cm]{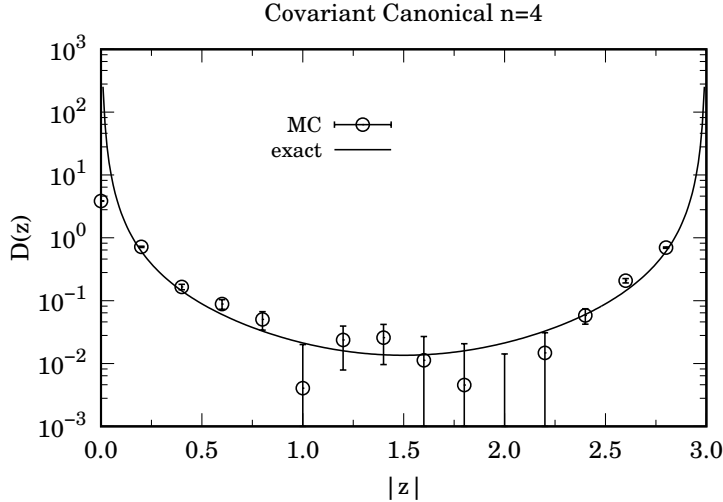}
\end{center}
\caption{Two-point function $D(z)$ of Eq. (\ref{eq:tp}), for a free 
real scalar field subject to canonical quantization with a self-interaction potential 
density of the form $V(\phi)=\frac{1}{2}m^2\phi^2$ in Eq. (\ref{eq:c-action}) with 
$m=1, N=15, L=3, a=L/N=0.2$. We compare with the analytic exact expression of Eq. 
(\ref{eq:tpt}) with an infrared cutoff of $irc=2$. A logarithmic scale is used on the 
$y$-axis.}
\label{fig:tp-c-N15L3m1}
\end{figure}

For a massive free-theory, $V(\phi)=\frac{1}{2}m^2\phi^2$, in canonical quantization 
(\ref{eq:c-action}) with $m=1, N=15, L=3, a=L/N=0.2$ we obtained the result shown in 
Fig. \ref{fig:tp-c-N15L3m1} where we compare the MC results with the exact expression 
of Eq. (\ref{eq:tpt}) with an infrared cutoff of $irc=2$ which is sufficient for an 
accuracy of $10^{-3}$. The run was $M=10^6$ MC steps long. The figure shows good 
agreement between the MC and the exact expression except at the origin due to the 
space-time discretization.

\begin{figure}[htbp]
\begin{center}
\includegraphics[width=10cm]{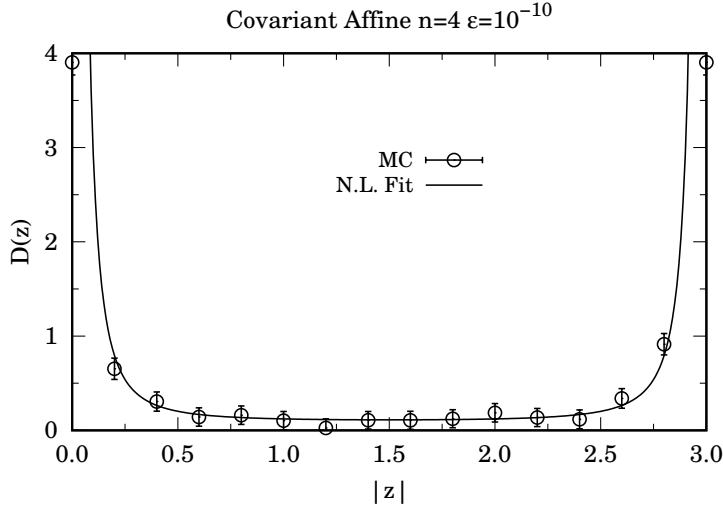}
\end{center}
\caption{Two-point function $D(z)$ of Eq. (\ref{eq:tp}), for a free 
real scalar field subject to affine quantization with term A and a self-interaction 
potential density of the form $V(\phi)=\frac{1}{2}m^2\phi^2$ in Eq. (\ref{eq:a-action}) 
with $m=1, N=15, L=3, a=L/N=0.2$, and $\epsilon=10^{-10}$. Also shown is the result of 
a non-linear fit of the data (except the first point at $|z|=0$) with the function 
$D_{m_D}(z)$ where $D_{m_D}$ is the two-point function of a free field of mass $m_D$ 
of Eq. (\ref{eq:tpt}) with an $irc=2$, taking $m_D$ as the only fit parameter.}
\label{fig:tp-a-N15L3m1}
\end{figure}

For a free massive theory $V(\phi)=\frac{1}{2}m^2\phi^2$ in affine quantization 
(\ref{eq:a-action}) using term A, the self-interaction is a double well with a spike 
barrier at $\phi=0$. We tuned the width of the displacement, $\delta$ in Eq. 
(\ref{eq:move}), so that the random walk in the $\phi(x)$ will sample the probability 
distribution $\exp(-S[\phi])$ most efficiently, with short equilibration times. In Fig. 
\ref{fig:tp-a-N15L3m1} we show the result for a free real scalar field subject to 
affine quantization with a total self-interaction of the form 
${\cal V}(\phi)=\frac{1}{2}m^2\phi^2+\frac{3}{8}a^{-2s}/(\phi^2+\epsilon)$ 
with $m=1, N=15, L=3, a=L/N=0.2$, and $\epsilon=10^{-10}$ after cutting the first 
equilibration MC steps of a run made of $M=2.5\times 10^6$ steps. During the 
simulations we also calculated the renormalized mass $m_R$ and the renormalized 
coupling constant $g_R$ \cite{Fantoni2020}. As we can see from the figure the symmetry 
$z \to L-z$ of the two-point function is preserved within the errorbars. 
The minima of the classical ${\cal V}$ is at $\phi=\pm\Phi$ with 
$\Phi^2=-\epsilon+\sqrt{3}/(2a^3m)$ which diverges in the continuum limit $a\to 0$ 
(this of course does not happen in the harmonic oscillator case \cite{Gouba2020} which 
is independent of the lattice spacing). Moreover the minimum of the action 
$L^{s+1}m(\sqrt{3}-m\epsilon a^s)/2a^s$ also diverges, both in the continuum limit at 
finite volume ($ma \to 0$) and in the infinite volume limit at fixed lattice spacing 
($mL \to \infty$) (this also happen for the affine harmonic oscillator \cite{Gouba2020} 
which has a well defined zero temperature limit). The corresponding contribution to the 
vacuum expectation only occurs together with the normalization constant in front of the 
path integral and drops out in quantities of physical interest (as long as the system 
is not placed in a curved geometry, i.e. in a gravitational field - there, the 
cosmological constant does have physical significance)

The symmetry $\phi\to-\phi$ is broken in the simulations (see Appendix \ref{app:Phi}) 
and as a result $\langle\phi(x)\rangle$ is different from zero. The action 
$S=\bar{K}+\bar{V}$ where $\bar{K}$ is the kinetic term and $\bar{V}$ the total 
potential term. Imagine now that we are in a configuration where all the $N^n$ 
components, $\phi(x)$, are around $+\Phi$. In order to start migrating one single $x'$ 
component, $\phi(x')$, around the other minimum at $-\Phi$ will have no cost in the 
potential, $\Delta\bar{V}\approx 0$, but it will have a big cost in the kinetic term 
between ``neighboring'' $x$, resulting in a big $\Delta\bar{K}$ (as long as the 
distance between the two minima, $2\Phi$, which diverges in the continuum limit, is 
large). As a consequence $\exp(-\Delta S)$ will be very small and the move will be 
almost surely rejected according to the Metropolis rule. Moreover, once the system 
reaches the phase with all $\phi(x)$ in one of the minima, it is very unlikely that a 
single $\phi(x')$ will move to the other minimum but it cannot be excluded, in 
principle. If this happens one has a situation where the field is around $+\Phi$ at all 
$x$ except at $x'$ where it is around $-\Phi$. But we can easily see that now it would 
be statistically favorable for the single field on the left to rejoin the fields on the 
right other then all the fields on the right join the field on the left. Exactly the 
same holds for affine quantization (\ref{eq:a-action}) using term B, since due to the 
kinetic energy term in the action the fields at neighboring points tend to assume 
similar values. On the other hand this would not 
hold for an {\sl ultralocal} \cite{Klauder2020b} theory where we could have the field 
visiting both wells at $\pm\Phi$ but only at not ``neighboring'' times, resulting in a 
vanishing $\langle\phi(x)\rangle$. Apart from this the shape of the two-point function 
is qualitatively similar to the one of the {\sl covariant} case of Eq. 
(\ref{eq:a-action}). In addition in a covariant {\sl complex} field one could go 
``slowly'' ``around'' the ``mountain'' at $\phi=0$  with no need of ``jumps''. 
 
For our choice of the parameters we have $\Phi\approx 10.404$ with $\Phi^2\approx 
108.253$. The results in Fig. \ref{fig:tp-a-N15L3m1} indicate that the quantization 
increases this number by about 10\%. The minimum of $D(z)$ is reached around $|z|=L/2$. 
The two-point function is qualitatively similar to the one of the free field. This is 
supported by recent results on a one dimensional harmonic oscillator treated with 
affine quantization \cite{Gouba2020} where it is shown that the eigenvalues are still 
equally spaced. A non-linear fit of the MC data (removing the first point at $|z|=0$) 
with the function $D_{m_D}(z)$ where $D_{m_D}$ is the two-point function of a free 
field of mass $m_D$ of Eq. (\ref{eq:tpt}) with an $irc=2$, taking $m_D$ as the only fit 
parameter, gives $m_D\approx 0.9$. The result of the fit is also shown in Fig. 
\ref{fig:tp-a-N15L3m1}.

For a free real scalar field subject to affine quantization with term A, in $n=4$ 
space-time dimensions in a volume $3^4$ with a regularization parameter 
$\epsilon=10^{-10}$, we studied the continuum limit, $N\to\infty$, (by choosing values 
lower of 15) and the dependence on the bare mass $m$, of the five quantities 
$m_R, g_R, \langle\phi(x)\rangle^2, m_D$, and $D(0)$. The results are shown in Table 
\ref{tab:study}. From the table we see how moving towards the continuum limit 
$m_D\approx m$ but $m_R$ becomes small due to the fact that when the field picks up an 
expectation value, the Fourier transform of the field $\widetilde{\phi}(0)$ picks up a 
contribution proportional to the volume of the box. Moreover, for the same reason, 
$g_R\approx 2$. The Table also shows the value of $\Phi^2$ and of 
$\langle\phi(x)\rangle^2$ to be compared. We see that the second is always larger than 
the first one by a percentage increasing with increasing $m$ and with increasing $a$. 
The value of $D(0)$ is increasing with a decrease of the lattice spacing $a$, signaling 
a divergence in the continuum limit.

\begin{table}[htbp]
\caption{We determined, for a free real scalar field subject to affine quantization 
with term A, in $n=4$ space-time dimensions, the dependence of $m_R, g_R, m_D$, and 
$D(0)$ on the number of one dimensional discretization points $N$ and the bare mass 
$m$ at $L=3$ with a regularization parameter $\epsilon=10^{-10}$. The runs were 
$M=5\times 10^6$ MC steps long. The value of $\Phi^2$ and of $\langle\phi(x)\rangle^2$ 
are also shown for comparison.} 
\label{tab:study}
\vspace{.5cm}
%\begin{ruledtabular}
{\footnotesize 
\begin{tabular}{||c|c||cccccc||}
\hline
\hline
$N$ & $m$ & $m_R$ & $g_R$ & $\Phi^2$ & $\langle\phi(x)\rangle^2$ & $m_D$ & $D(0)$\\
\hline
\hline 
\multirow{3}{*}{15} & 1 & 0.0122(3) & 1.9979(1) & 108.2 & 120.6(1) & 0.934 & 3.69(6)\\
%\hline
 & 2 & 0.00646(4) & 1.99983(3) & 54.13 & 65.7(1) & 1.785 & 3.32(6)\\
%\hline
 & 3 & 0.0186(6) & 1.99925(8) & 36.08 & 45.85(7) & 3.009 & 2.97(6)\\
\hline
\multirow{3}{*}{12} & 1 & 0.01053(5) & 1.99958(5) & 55.43 & 63.25(8) & 0.302 & 2.38(5)\\
%\hline
 & 2 & 0.00967(9) & 1.99992(2) & 27.71 & 34.54(5) & 2.467 & 2.00(5)\\
%\hline
 & 3 & 0.0095(1) & 1.99905(8) & 18.47 & 24.00(4) & 5.483 & 1.66(5)\\
\hline
\multirow{3}{*}{10} & 1 & 0.01417(4) & 1.999464(4) & 32.07 & 37.46(5) & 0.587 & 1.58(3)\\
%\hline
 & 2 & 0.0124(1) & 1.99995(1) & 16.04 & 20.43(3) & 3.789 & 1.29(3)\\
%\hline
 & 3 & 0.0119(2) & 1.99996(1) & 10.69 & 14.03(2) & 5.647 & 1.02(3)\\
\hline
\hline
\end{tabular}
}
%\end{ruledtabular}
\end{table}

Summarizing, the two-point function for $\phi - \langle\phi\rangle$ 
looks similar to the two-point function of a free field with mass $m_D$. In other 
words, the correlation length of the affine quantum field theory is $m/m_D$ times the 
Compton wavelength of the canonical quantum theory of the free scalar field. Our 
results seem to suggest that, going towards the continuum, the affine model is 
approaching a free field with the same bare mass.

The value of $m_D$ is not easy to understand, however. If the action is treated at the 
classical level, small deviations from the minimum are determined by the curvature of 
the total potential, $m_c^2 = d^2{\cal V}/d\phi^2$ at $\phi = \Phi$.
The mass term contributes $m^2$ and the ``3/8'' term yields a contribution that is 3 
times larger. For $\epsilon = 0$, the mass relevant for the relation between frequency 
of the waves and wavelength is: $m_c = 2 m$ independently of $a$.
 
\begin{figure}[htbp]
\begin{center}
\includegraphics[width=10cm]{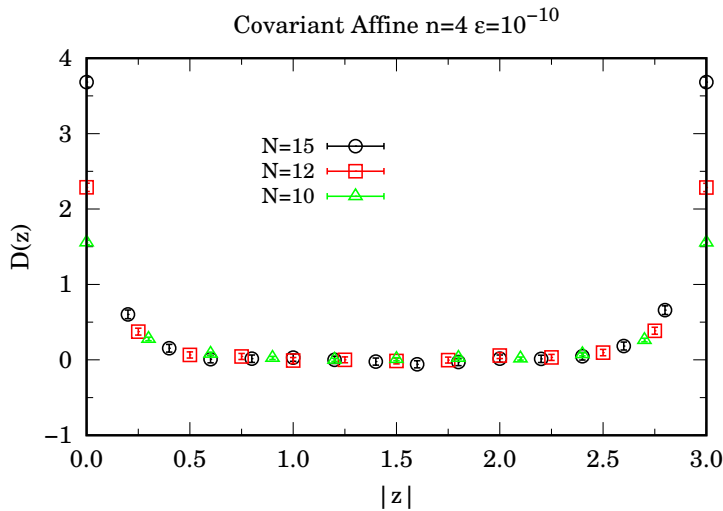}
\end{center}
\caption{(color online) Two-point function $D(z)$ of Eq. (\ref{eq:tp}) for a free 
real scalar field subject to affine quantization with a self-interaction potential 
density of the form $V(\phi)=\frac{1}{2}m^2\phi^2$ in Eq. (\ref{eq:a-action}) with 
$m=1, L=3, \epsilon=10^{-10}$ and increasing $N=10,12,15$.}
\label{fig:tp-a}
\end{figure}

In Fig. \ref{fig:tp-a} we show $D(z)$ as obtained for $m=1$ 
($L=3, \epsilon=10^{-10}$) and three choices of $N$, in the long simulations of the 
Table \ref{tab:study}. One can then see the approach to the continuum of the two-point 
function of the affine model.

%

%%%%%%%%%%%%%%%%%%%%%%%%%%%%%%%%%%%%%%%%%%%%%%%%%%%%%%%%%%%%%%%%%%%%%%%%%%%%%%
\section{Conclusions} 
%%%%%%%%%%%%%%%%%%%%%%%%%%%%%%%%%%%%%%%%%%%%%%%%%%%%%%%%%%%%%%%%%%%%%%%%%%%%%%
\label{sec:conclusions}

In a recent work \citep{Fantoni2020a} we studied the case of a non-renormalizable 
$(\phi^4)_4$ canonical theory (where the self-interaction potential is 
$V(\phi)=g\phi^4$) in four space-time dimensions and proved through MC that the theory 
becomes renormalizable if one treats the field through affine quantization. 

In the present work we observed that for $g=0$ the simplest question to ask was: Does 
the affine system describe particles as for the canonical one? If so, do they interact 
with one another? 
 
We tried to answer these question by looking at the two-point function. What we proved 
through our MC analysis was that the affine case with $g=0$ has to be considered like a 
``sort'' of free-theory of ``quasiparticles'' (in the sense of Lev Landau in his theory 
for Fermi liquids) where the ``3/8'' term just offers itself like a sort of 
``collective excitation'' term. In this case the $\phi\to-\phi$ symmetry is broken and 
the field acquires a non-zero vacuum expectation. The two-point function nonetheless 
has all the same features as those of a free scalar field of similar mass, in the 
continuum limit.

One shortcoming of the affine formulation of the field theory is the divergence (in the 
continuum) of the vacuum expectation value of the field which generates the 
disconnected contribution to the Green's functions. The path integral is fully 
determined by the local properties of the field that enter through the action. The 
expectation value of the field does not represent a local property of the field. We 
cannot imagine how one could possibly get rid of it. In the Standard Model, however, 
one of the crucial properties of the Higgs fields is that they pick up a vacuum 
expectation value $v$. The masses of the W- and Z-bosons as well as those of the 
leptons and quarks are proportional to $v$. In order to remedy to this drawback one 
should perform the following scaling $\phi\to a^{-s/2}\phi$ (together with $g\to a^sg$ 
in a possible interaction term of the form $g\phi^4$) which would bring about an 
additional factor $a^{-s}$ multiplying the action. This scaling proved successful in 
our forthcoming work on the affine quantization of a Higgs complex scalar field 
\cite{Fantoni2021}.

The present paper is wanted to confirm that both canonical and affine procedures lead 
to desired and expected behavior for quadratic potential terms. A later paper 
\cite{Fantoni2021} will be designed to deal with quartic potential terms with canonical 
and affine procedures.

\appendix
%%%%%%%%%%%%%%%%%%%%%%%%%%%%%%%%%%%%%%%%%%%%%%%%%%%%%%%%%%%%%%%%%%%%%%%%%%%%%%
\section{Field configurations in the vicinity of the two degenerate minima in the 
affine version}
%%%%%%%%%%%%%%%%%%%%%%%%%%%%%%%%%%%%%%%%%%%%%%%%%%%%%%%%%%%%%%%%%%%%%%%%%%%%%%
\label{app:Phi}
Classically, the affine version of the free Hamiltonian has two degenerate minima,  
$\phi = \pm \Phi$. If the path integral is dominated by those field configurations that 
are located in the vicinity of one of these everywhere on the entire lattice or in the 
vicinity of the other, then it consists of two equal pieces
\bq \nonumber
Z   &=&\int{\cal D}\phi\,\exp(-S[\phi]),\\\nonumber
Z_+ &=&\int{\cal D}\phi\,\exp(-S[\phi]),\;\;\mbox{integral only over}\;\;\phi(x)\approx\Phi,\\\nonumber
Z_- &=&\int{\cal D}\phi\,\exp(-S[\phi]),\;\;\mbox{integral only over}\;\;\phi(x)\approx-\Phi,
\eq
and $Z_+ = Z_-$. Under a broken symmetry $\phi\to-\phi$ one would get either 
$Z \approx Z_+$ or $Z \approx Z_-$. This has to be expected in the present case of a 
real field since in order to move the field $\phi(x)$ at a single $x$ from around 
$\Phi$ to around $-\Phi$ in the MC path integral one has to overcome a large kinetic 
cost. This is not true for a complex field where one can 
go ``slowly'' ``around'' the ``mountain'' at $\phi=0$.

The expectation value of the field
\bq \nonumber
\langle\phi(x)\rangle   &=& \int{\cal D}\phi\,\phi(x)\exp(-S[\phi])/Z,\\\nonumber
\langle\phi(x)\rangle_+ &=& \int{\cal D}\phi\,\phi(x)\exp(-S[\phi])/Z_+,\;\;\mbox{over}\;\;\phi(x)\approx\Phi\\\nonumber
\langle\phi(x)\rangle_- &=& \int{\cal D}\phi\,\phi(x)\exp(-S[\phi])/Z_-,\;\;\mbox{over}\;\;\phi(x)\approx-\Phi,
\eq
with $\langle\phi(x)\rangle_+ \approx \Phi, \langle\phi(x)\rangle_- \approx -\Phi$, and
under the broken symmetry, 
$\langle\phi(x)\rangle \approx \langle\phi(x)\rangle_\pm\approx\pm\Phi$
where the simulation, starting from $\phi=0$, will choose among the two different cases 
just after the first equilibration steps.

For the two-point function
\bq \nonumber
D_+(x-y)&=&\int{\cal D}\phi\,\phi(x)\phi(y)\exp(-S[\phi])/Z_+-\langle\phi(x)\rangle_+^2,\;\;\mbox{over}\;\;\phi(x)\approx\Phi,\\\nonumber
D_-(x-y)&=&\int{\cal D}\phi\,\phi(x)\phi(y)\exp(-S[\phi])/Z_+-\langle\phi(x)\rangle_-^2,\;\;\mbox{over}\;\;\phi(x)\approx-\Phi,
\eq
so that $D_+(z)\approx 0, D_-(z)\approx 0$, and $D(z)\approx D_\pm(z)\approx 0$. 

Moreover one can see how in the broken symmetry configuration in which $\phi^2(x)
\approx\Phi^2\sim a^{-3}$, the ``3/8'' term in the Hamiltonian density is also of 
the same order in the continuum limit $a\to 0$. This will lead to a convergent 
two-point function for $\phi-\langle\phi\rangle$ in the continuum limit.
 
%%%%%%%%%%%%%%%%%%%%%%%%%%%%%%%%%%%%%%%%%%%%%%%%%%%%%%%%%%%%%%%%%%%%%%%%%%%%%% 
\begin{acknowledgments}
Many thanks to Heinrich Leutwyler for his suggestions, comments, and someone who 
proposed the canonical and affine programs with regard to the required features 
including what to examine and what to expect, which has influenced our program and its 
results, and led to many positive results and highlighted the expected canonical and 
affine differentiation.
\end{acknowledgments} 
%%%%%%%%%%%%%%%%%%%%%%%%%%%%%%%%%%%%%%%%%%%%%%%%%%%%%%%%%%%%%%%%%%%%%%%%%%%%%%
%\bibliographystyle{}
%\bibliography{higgs}

%apsrev4-2.bst 2019-01-14 (MD) hand-edited version of apsrev4-1.bst
%Control: key (0)
%Control: author (8) initials jnrlst
%Control: editor formatted (1) identically to author
%Control: production of article title (0) allowed
%Control: page (0) single
%Control: year (1) truncated
%Control: production of eprint (0) enabled
%

%%%%%%%%%%%%%%%%%%%%%%%%%%%%%%%%%%%%%%%%%%%%%%%%%%%%%%%%%%%%%%%%%%%%%%%%%%%%%%
%%%%%%%%%%%%%%%%%%%%%%%%%%%%%%%%%%%%%%%%%%%%%%%%%%%%%%%%%%%%%%%%%%%%%%%%%%%%%%
%%%%%%%%%%%%%%%%%%%%%%%%%%%%%%%%%%%%%%%%%%%%%%%%%%%%%%%%%%%%%%%%%%%%%%%%%%%%%%
\end{document}